\documentstyle[aps]{revtex}
\begin{document}
\title{On the classical hydrodynamic limit of quantum field theories}
\author{A.M. Lisewski}
\address{Max-Planck-Institut f\"ur Astrophysik,
Karl-Schwarzschild-Str. 1, 85740 Garching, Germany}
\date{\today}
\maketitle
\begin{abstract}
We discuss the transition from a quantum to a classical domain for
a model where a separation into environment and system is explicitely not given.
Utilizing the coarse graining procedure for free quantum fields we
also apply the projection method and the Hamiltonian principle to study
possible cases of emergent classicality. General conditions
for classical dynamics are given. Eventually, they lead to the equations of motion for a
perfect classical fluid.
\end{abstract}

\section{Introduction}
The transition from a quantum mechanical
description to a classical one for systems with many degrees of
freedom is still an active field of research. 
An interesting issue among it is the emergence
of classical hydrodynamical equations out of a many-particle quantum
system. It is commonly believed that one has to accomplish 
two main tasks in order to warrant this emergence.
One is to show that entering the
classical domain implies that a small set of variables (local
densities of particle number, momentum and energy) experience
{\it decoherence}, where this notion qualitatively means that
quantum superpositions that give rise to non-classical phenomena 
(e.g., non-locality) disappear. 
The other is that those variables turn out
to obeye classical, hydrodynamic equations.
Among different approaches to this problem, Gell-Mann and Hartle
\cite{gmh93} present the idea of how local densities may become classical when
integrated over small volumes of space (method of coarse graining). As a consequence, the coarse grained variables
are acompanied by dissipation and fluctuations. Furthermore, Gell-Mann and
Hartle describe how decoherence, dissipation and noise may interact
in such a manner that classicality becomes present.  
Physically, this approach should be appealing in systems where it 
is practically impossible to distinguish between a system and its environment,
like in gases or in fluids. However, until now there were hardly
any physically interesting models on which one could explicitely
show this way to emergent classiality when a clear system-environment-separation is missing.\\
In this work we discuss the possiblity of emergent classicality exactly
for these kind of systems. First of all, we have to point out
what is actually meant by the notion 'classical'. 
The underlying assumption is that the classical realm is characterized 
by two general properties. This is
determinism and locality in space as well as in time.
Thus we want to find out, under which conditions
these properties become valid.
We start from a non-relativistic and non-interacting quantum field
theory. Using Zwanzig's projection method, we give an evolution equation
for the expectation values of the coarse grained fields.
The expectation values themselves are represented in a fixed $ N$-particle
Hilbertspace. Then we apply Hamilton's principle to 
express this evolution through equations of motion for local densities.
It turns out that under certain conditions determinism and locality in
time are consistent with the resuluts from the Hamiltonian principle. In fact, this condition is
the existence of certain values of the {\it averaging length},
$l_{\rm av}$,
which parametrizes the roughness of coarse graining.\\
The other classical property, locality in space, needs further
discussion. In this context we discuss the problem of {\it nodal regions}
. Their existence leads inevitably to additional quantization
rules. These rules have no classical meaning and hence we want
them to become trivial in the classical regime. This turns out to happen
when the coarse-grained density is positive everywhere.
As next we show that non-local quantum effects become
unimportant when the 
{\it quantum potential} in the equations of motion fulfills an inequality. 
This inequality gives an upper bound for the contribution of quantum
forces and relates them to microscopic fluctuations and dissipation.
Again this relation sets a condition for $ l_{\rm av}$.
\\
If all these conditions become consistent we find
the classical hydrodynamical equations for a perfect fluid, namely the
mass continuity equation and the Euler equation.
Finally, we give heuristic arguments of how $ l_{\rm av}$ can be 
interpreted as the temperature of the classical macrosystem.

\section{Evolution equations for coarse-grained fields}

We briefly summerize the
formal approach commomnly known as the {\it projection method}
which goes back to the original work of Zwanzig (see \cite{rm96}, and
the references therein). 
It provides us with an evolution equation for the relevant part of the
quantum system's statistical operator $ \rho$. The subspace
containing the relevant states is represented by a projector
{$\cal P$}. Of course, the realization of {$\cal P$} is
motivated by certain physical considerations but for the projection
method itself the choice of {$\cal P$} is arbitrary.
Let $ {\cal H}$ be the Hamiltonian of the system and $ A$ an
operator. Then the evolution equation for the expectation value of 
the relevant part of $ A$, that is
$ \langle A \rangle(t)  = {\rm tr} [ \rho(t) {\cal P} A ]$, reads as
\begin{equation}
\label{zwanzig1}
i \frac{\partial}{\partial t} \langle A \rangle(t) - \langle {\cal P L}
A \rangle(t) + i \int\limits_{0}^{t} {\rm d}\tau \langle {\cal P L Q}
\exp(i {\cal L} \tau) {\cal Q L P} A \rangle(t - \tau) =
\zeta_{A}(t) \,,
\end{equation} 
where $ {\cal Q} = 1 - {\cal P}$ and $ {\cal L} \,\cdot \equiv [{\cal H}, \,\cdot\,]$ denotes
the Liouvillian (We set $ \hbar = 1$). The term on the right hand side
represents the influence of the irrelevant states at initial time,
that are states
orthogonal to ${\cal P} \rho(0)$. It is
\begin{equation}
\zeta_{A}(t) = - {\rm tr} [ \rho(0) {\cal Q} \exp(i {\cal L} t) {\cal
Q} {\cal L} A] \,.
\end{equation}
We apply this result for non-interacting quantum theory of 
fields in the non-relativistic limit ($ c \rightarrow\infty$). Thus $
A$ is now considered to be a field operator in the 
Heisenberg picture, $ \hat{a}(x)$, such that $ \hat{a}(x;t) = \exp(i
{\cal H} t) \,\hat{a}(x)\, \exp(-i {\cal H} t)$ satisfies 
the Schr\"odinger equation.
The aim is to then represent equation (\ref{zwanzig1}) in the $
N$-particle Hilbert space $ {\textsf H}^{N}$ and to use it later for
the construction of
effective hydrodynamic equations. However, these equations will
not be classical a priori. Only if we find certain conditions for a 
classical description then hydrodynamic equations will possibly emerge.\\
Let $ a(1, ..., N; t) := a(x_1, ..., x_N; t) = 
{\rm tr}_{\scriptsize{\textsf H}^N}[\rho(t){\cal P} \hat{a}(x;0)]$ be the
relevant part of the $ N$-particle wave function represented in
position space. The evolution equation for $ a(1,...,N; t)$ can then
be written as
\begin{eqnarray}
\label{Nzwanzig}
\nonumber
i \frac{\partial}{{\partial} t} a(1,...,N; t) + \int  {\rm d}1' ... {\rm d}N'
 \, & H^{\cal P} & (1',...,N';t) \,a(1',...,N';t) = \\ 
&  & \zeta_{\hat a}(1,...,N; t)  - i\int {\rm d}1' ... {\rm d}N'  
\int\limits_{0}^{t} {\rm d}\tau \,G^{\cal P}(1,...,N,1',...N';\tau) \,a(1',...,N';
t - \tau).
\end{eqnarray}
Herein $ H^{\cal P}$ and $ G^{\cal P}$ are the integral kernels 
representing the operators ${\cal H^P}a = {\cal P H}a$ and of 
\begin{equation}
{\cal G^P}a =
\int\limits_{0}^{t} {\rm d}\tau {\cal P
H Q} \,\exp(i{\cal H} \tau)\, {\cal Q H P}a(t - \tau) \,.
\end{equation}
The fluctuation term $ \zeta_{\hat a}(1,...,N; t)$ is also written
as an integral operator, viz.
\begin{equation}
\nonumber
\zeta_{\hat a}(1,...,N; t) = \int {\rm d}1' ... {\rm d}N' \,F(1',...,N',1,...,N; t)
\,a(1',...,N'; 0)\,,
\end{equation}
where $ a(1',...,N'; 0)$ is the original, i.e. the 'unsmeared', wavefunction at initial time.
A similar equation for $ a^\dagger(1,...,N; t) = 
{\rm tr}_{\scriptsize{\textsf H}^N}[\rho(t){\cal P}
\hat{a}^\dagger(x;0)]$ is obtained by taking the complex conjugate of 
equation (\ref{Nzwanzig}).\\
The actual realization of the operator
kernels depends on the choice of the projector ${\cal P}$.
A natural choice of $ {\cal P}$ is that of covering the whole space
with small cubes of volume $ l_{\rm av}^3$, taking the average 
on each of these cubes such that the set of all average values defines
the coarse grained function. Therefore the spatial components of the
(coarse grained) wave functions $ a(1,...,N;t)$ are defined on a cubic lattice $
\Sigma_{l_{\rm av}}$ with
grid size $ l_{\rm av}$. In that case the integral kernel $ P(x, x')$ of ${\cal
P}$ is the characteristic function of the set $ [x - l_{\rm av}/2, x +
l_{\rm av}/2]^3 \subset {\sf R}^3$ for a given $ x \in \Sigma_{l_{\rm av}}$. For other
choices of ${\cal P}$ we require that for any $ x \in \Sigma_{l_{\rm
av}}$ the support of $ P(x, \cdot)$ is  basically a set of
volume $ l_{\rm av}^3$ such that for all $ x'$ with $ |x - x'| >
l_{\rm av}$ the corresponding value of $ P(x,x')$ is suppressed 
at least exponentially. This requirement reflects the physical assumption that
all wavelengths shorter than $ l_{\rm av}$ should not contribute to the
physical description on scales much larger than $ l_{\rm av}$.
Independently of this particular choice we can state that 
if $ l_{\rm av} \ll l_{\rm obs}$, where $ l_{\rm obs}$ is the minimum
length scale on which macroscopic functions significantly vary,
then the grid $ \Sigma_{l_{\rm av}}$ is taken to be the continuum $
{\sf R}^3$ again 
\footnote{In general, the operator $ {\cal P}$ satisfying the above condition
will not be a projector anymore, i.e. $ {\cal P}^2 \neq {\cal P}$,
but on the macroscopic level the deviation $|{\cal P}^2 - {\cal P}|$
will be only of the order $(l_{\rm av}/l_{\rm obs})$ \cite{a98}.}. 
Until now, we have not given any physical motivation
or mathematical indication that confirms the inequality $ l_{\rm av} \ll l_{\rm
obs}$. A possible confirmation of this assumtion will be given later,
when the emergence of classicality will be discussed.\\
As an obvious realization of
${\cal P}$ we take Gaussian distribution for $ P(x,x')$,
\begin{equation}
P(x,x') = \int {\rm d}k \,\exp[-i k (x - x')]\,\exp[- k^2l_{\rm av}^2/2]
\,,
\end{equation}
where the short form $ {\rm d}k$ stands for $ (2 \pi)^{-d/2} \, {\rm d} k_1 ... {\rm d}k_d$.
This expression is  generalized for the $ N$-particle state via
\begin{equation}
\label{NP}
P(x_1,...x_N,x_1'...x'_N) = \int {\rm d}k_1...{\rm d}k_N 
\,\exp[-i \sum_{j = 1}^{N}k_j (x_j - x_j')]
\,\exp[-\sum_{j = 1}^{N} k^2_jl_{\rm av}^2/2] \,.
\end{equation}
Using $ P(k_1,...,k_N) = \exp[-1/4 \sum_{j = 1}^{N} k^2_jl_{\rm av}^2]$ we
can write down 
the expressions for $ G^{\cal  P}$ and $ H^{\cal P}$, which are
\begin{eqnarray}
\nonumber
F^{\cal P}(x_1,...,x_N, x'_1,...,x'_N; t) = \int {\rm d}k_1 ... {\rm d}k_N\, 
\omega_{k_1,...,k_N} P(k_1,...,k_N) [1 - P(k_1,...,k_N)]^2  \times\\
 \times\,\exp[-i\sum\limits_{j = 1}^{N} k_j (x_j -
x'_j) - i \,\omega_{k_1,...,k_N} t]
\end{eqnarray}
\begin{eqnarray}
\nonumber
G^{\cal P}(x_1,...,x_N, x'_1,...,x'_N; t) = \int {\rm d}k_1
... {\rm d}k_N\,
\omega_{k_1,...,k_N}^2 [P(k_1,...,k_N)
- P(k_1,...,k_N)^2 ]^2 \times\\
 \times\,\exp[-i\sum\limits_{j = 1}^{N} k_j (x_j -
x'_j) - i \,\omega_{k_1,...,k_N} t]
\end{eqnarray}
\begin{eqnarray}
H^{\cal P}(x_1,...,x_N, x'_1,...,x'_N; t) = \int {\rm d}k_1
... {\rm d}k_N\,
\omega_{k_1,...,k_N} P(k_1,...,k_N)^2
 \,\exp[-i\sum\limits_{j = 1}^{N} k_j (x_j -
x'_j)].
\end{eqnarray}
Herein, $ \omega_{k_1,...,k_N}$ denotes the eigenvalues of $ {\cal H}$
in Fourier-space.
These expressions seem quite complicated, but can be simplified significantly
by expanding $ P(k_1,...,k_N)$ to the lowest order in $ l_{\rm av}$, 
when it is $ k^2 l_{\rm av}^{2} \ll 1$ which is equivalent to 
$ l_{\rm av} \ll  l_{\rm obs}$.
This gives for the Gaussian kernel of $ {\cal P}$
\begin{equation}
\label{HPapprox}
H^{\cal P}(x_1,...,x_N, x'_1,...,x'_N; t) = \int {\rm d}k_1
... {\rm d}k_N\, \omega_{k_1,...,k_N}
 \,\exp[-i\sum\limits_{j = 1}^{N} k_j (x_j -
x'_j)]\,,
\end{equation}
\begin{equation}
\label{GPapprox}
G^{\cal P}(x_1,...,x_N, x'_1,...,x'_N; t) = \frac{l_{\rm av}^8}{64}\,\int {\rm d}k_1
... {\rm d}k_N\,(k_1 + ... + k_N)^4 \omega_{k_1,...,k_N}^2\,\exp[-i\sum\limits_{j = 1}^{N} k_j (x_j -
x'_j) - i \,\omega_{k_1,...,k_N} t] \,,
\end{equation}
and for the kernel acting on the 'unsmeared' states
\begin{eqnarray}
\label{FPapprox}
\nonumber
F^{\cal P}(x_1,...,x_N, x'_1,...,x'_N; t) = \frac{l_{\rm
av}^4}{16}\,\int {\rm d}k_1 ... {\rm d}k_N\,& &(k_1 + ... + k_N)^2\,
\omega_{k_1,...,k_N}\\
& & \,P(k_1,...,k_N)\,\exp[-i\sum\limits_{j = 1}^{N} k_j (x_j -
x'_j) - i \,\omega_{k_1,...,k_N} t]\,.
\end{eqnarray}
Under this approximation, the averaging-length $ l_{\rm av}$
appears as a coupling constant determining the strength of the
dissipation and the fluctuation term \cite{a98}.

In appearence, equation (\ref{Nzwanzig}) shows uncommon features
in the classical regime. First there is  the non-local behavior in
time. Second, the fluctuation term makes it questionable
whether eqn. (\ref{Nzwanzig}) really describes a determinstic
evolution in time. Therefore, in order to assign classicality 
to the evolution equation (\ref{Nzwanzig}) we have to show
how these 'uncommon features' become unimportant. This is the 
issue of the following section and one of the main issues of this work.

\section{The classical equations}
\subsection{The formal equations}
For a hydrodynamical description of equation
(\ref{Nzwanzig}), we wish to find a representation by a small set of
variables -- in general those will be local densities of particle
number, momentum and energy -- such that these variables follow dynamic equations similar
to the fluid mechanical equations. This representation should be 
independent of the number of particles in the system, if only this 
number is large enough and does not strongly vary in time.
Then the resulting equations should be equivalent to the evolution equation
(\ref{Nzwanzig}). Consequently, we expect that the hydrodynamical
modes are subject to dissipation and noise.\\
In the remarkable work of Holzwarth and Sch\"utte \cite{hs78} a
 fluid-dynamical description of a multiple particle quantum system is given. Using a variational
scheme, it generalizes earlier attempts by considering also
two-particle correlations. This generalization leads to a velocity
field with non-vanishing vorticity being a direct consequence of
two-particle correlations.\\
We adopt this idea for the relevant wave function $ a(1,...,N;t)$ and
choose the following ansatz for $ a(1,...,N;t)$,
\begin{equation}
 a(1,...,N;t) = \phi(1,...,N;t) \,\exp[i \,m \,S(1,...,N;t)]\,,
\end{equation}
with the real phase $ S$ including a two-particle term of the form
\begin{equation}
S(1,...,N;t) = \sum\limits_{i = 1}^{N} \varphi(i;t) \,+\, \frac{1}{2} 
\sum\limits_{i \neq j} \mu(i;t) \mu(j;t) \,.
\end{equation}
Using this representation of $ a$ we construct a Lagrangian functional
$ L$ out of equation ({\ref{Nzwanzig})
\begin{equation}
\label{lagr}
L = \langle a| \left(\,i\, \partial_{t}|a\rangle - {\cal H^P} |a\rangle - {\cal
G^P} |a\rangle +
\zeta_{\hat a} \right)\,.
\end{equation}
Denoting  $ (\langle a|{\cal G^P} |a\rangle - \langle a|\zeta_{\hat a}) $ by $ E_{\cal
P}$ and making use of norm conservation, we have 
\begin{equation}
L = -m \int {\rm d}1 ... {\rm d}N \, \phi^2(1,...,N;t) \,\dot{
S}(1,...,N;t) - \frac{m}{2} \int {\rm d}1...{\rm d}N \,
\phi^2(1,...,N;t) \, \sum_{i = 1}^N \nabla_iS \cdot \nabla_iS - E_{\rm qm} 
- E_{\cal P}
\end{equation}
with
\begin{equation}
E_{\rm qm} = \frac{m}{2}\int {\rm d}1 ... {\rm d}N \,\phi(1,...,N;t) \, \sum_{i=1}^N
\Delta_i \phi(1,...,N;t) \,.
\end{equation}
Introducing the one-, two- and three-particle moments $\rho^{(1)},
\rho^{(2)}$ and $ \rho^{(3)}$ by the formula
\begin{equation}
\rho^{(i)} = \frac{N!}{(N - i)!} \int {\rm d}(i + 1) ... {\rm d}N \, \phi^2(1,...,N;t)\,,
\end{equation}
where we denote $ \rho^{(1)}$ by $ \rho$ and the functions $ \lambda(1;t)$, $\kappa^2(1;t)$ by 
\begin{eqnarray}
\rho(1;t) \,\lambda(1;t) =  N\int{\rm d}2 ... {\rm d}N \,
\phi^2(1,...,N;t) \sum_{j > 1}^N \mu(j;t)
 =  \int \rho^{(2)}(1,2) \,\mu(2) \,{\rm d}2\,,
\end{eqnarray}
\begin{eqnarray}
\nonumber
\rho(1;t) \kappa^2(1;t) & = & N \int {\rm d}2 ... {\rm d}N
\,\phi^2(1,...,N;t) \left[ \sum_{j > 1}^N\mu(j;t) - \lambda(1;t)
\right]^2\\
& = & \int\rho^{(3)}(1,2,3) \mu(2)\,\mu(3) \,{\rm d}2 {\rm d}3 +
\int\rho^{(2})(1,2)\mu^2(2) \,{\rm d}2 - \rho(1) \lambda^2(1)\,,
\end{eqnarray}
we are able to write the Lagrangian (\ref{lagr}) in the following form
\begin{eqnarray}
\label{lagru}
\nonumber
L = -m\int{\rm d}1\,\rho(1;t) \,& & [\dot\varphi(1;t) + \lambda(1;t)
\dot\mu(1;t)]\\
& & -\frac{m}{2} \int{\rm d}1\,\rho(1;t) \,[(\nabla\varphi(1;t) + \lambda(1;t) 
\nabla\mu(1;t))^2 + \kappa^2(1;t) (\nabla\mu(1;t))^2] - E_{\rm qm} -
E_{\cal P} \,.
\end{eqnarray}
Therein, the function $ E_{\rm qm}$ only depends on the local density
$ \rho(1;t)$, whereas the function $ E_{\cal P}$ in general depends on $
\rho, \varphi, \mu$ and $ \kappa$. If we now identify the expression 
$ \nabla \varphi + \lambda \nabla\mu$ with a general velocity field $
u$, then $ L$ already has the form familar from fluid dynamics \cite{t53}.
It is remarkable that $ u$ has a non-vanishing vorticity $
\nabla\lambda \times \nabla\mu$.
The corresponding equations of motion follow from the Langragian derivatives 
\begin{equation}
[L]_\sigma = \frac{\delta L}{\delta \sigma} - \frac{{\rm d}}{{\rm d}
t} \frac{\delta L}{\delta \dot\sigma} \equiv 0\,, \qquad \sigma \in I
= \{\rho,
\varphi, \lambda, \mu, \kappa\} \,.
\end{equation}
They are
\begin{eqnarray}
\label{eqnm}
\left[L\right]_\varphi & = & \dot\rho + \nabla\cdot(\rho u) + 
\frac{1}{m} \frac{\delta}{\delta \varphi} E_{\cal P} = 0 \,,\\
\left[L\right]_\lambda & = & \rho (\dot\mu + u \cdot \nabla\mu) + 
\frac{1}{m} \frac{\delta}{\delta \lambda} E_{\cal P} = 0\,,\\
\left[L\right]_\mu & = &\dot{(\rho\lambda)} + \nabla \cdot (\rho\lambda u + \rho
\kappa^2 \nabla \mu) + \
\frac{1}{m} \frac{\delta}{\delta \mu} E_{\cal P} = 0\,,\\
\left[L\right]_\rho & = & \dot\varphi + \lambda\dot\mu + \frac{1}{2} u^2 
+ \frac{1}{2} \kappa^2 (\nabla\mu)^2 + 
\frac{1}{m} \frac{\delta}{\delta \rho} (E_{\rm qm} + E_{\cal P}) =
0\,,\\
\label{eqnme}
\left[L\right]_\kappa & = & \rho \kappa (\nabla \mu)^2 + 
\frac{1}{m} \frac{\delta}{\delta \kappa} E_{\cal P} = 0 \,.
\end{eqnarray}
The gradient of equation (26) together with the other equations of
motion leads to
\begin{eqnarray}
\label{euler1}
\dot{(\rho u)} + \nabla \cdot (\rho u \otimes u + \rho \kappa^2
\nabla\mu\otimes\nabla\mu) =
-\frac{1}{m}\left(\rho\nabla\frac{\kappa}{2 \rho}\frac{\delta}{\delta \kappa}
+\nabla \rho\frac{\delta}{\delta\rho} - \sum_{\sigma \in
I \backslash \{\kappa\}} (\nabla \sigma)\frac{\delta}{\delta
\sigma}\right) (E_{\rm qm} + E_{\cal P}) \,.
\end{eqnarray}
This equation looks very similar to the Euler equation except for the
right hand side, where additional terms appear. Those are the forces
originating from the coarse grained description and from the gradient of the
quantum potential $ \delta_\rho E_{\rm qm}$. It is important to note
that the value of the 'thermal pressure' $ P = \rho\kappa^2\nabla\mu
\otimes \nabla\mu$ is always positive. This holds independently of
the parity (odd or even) of $ S$, see \cite{hs78}.

\subsection{Constraints for classicality}
To establish classical behavior of eqn. (\ref{euler1}) it is necessary
to show its deterministic and local-in-time behavior as well as its
independence of any quantum effects. We utilize these guiding principles
in order formulate sufficient conditions for the emergence of macroscopic
classicality. Eventually, these conditions will specify our choice of the
averaging length, $ l_{\rm av}$.
\subsubsection{Determinism and locality in time}
At first, we discuss the important question of how to 'choose' a  value
of $ l_{\rm av}$ such that equation (\ref{euler1}) loses the 
non-local evolution in time and the non-deterministic nature
due to the terms in $ E_{\cal P}$.
We approach this problem by introducing a generaliziation of the 
variational scheme presented in the previous section. Until now,
$ l_{\rm av}$ was regarded as a free parameter in
our equations. It appears in the Lagrange functional (\ref{lagru})
and therefore we change the value of $ L$ whenever we change
$ l_{\rm av}$. Thus it is obvious to vary $ L$ also with respect
to $ l := l_{\rm av}$ and add the resulting equation to the sample
of equations (23) - (27). In doing so, the functional derivative
in $ [L]_{l}$ becomes an ordinary one and by virtue of equation
(27) we obtain
\begin{equation}
\label{lvar}
\frac{\partial}{\partial l} \int\,{\rm d}1\,\rho \frac{\delta}{\delta\rho} (E_{\rm qm} +
E_{\cal P}) = 
\frac{\partial}{\partial l} (E_{\rm qm} + E_{\cal P}) \,.
\end{equation}
The most trivial solution to this equation is the case when
$ (E_{\rm qm} + E_{\cal P})$ can be represented as
\begin{equation}
\label{trivsol}
(E_{\rm qm} + E_{\cal P}) = \int {\rm d}1 \,\rho^{1/2}
\epsilon[\lambda, \varphi, \kappa, \mu] \,\rho^{1/2} \,.
\end{equation}
with $ \epsilon$ being independent of $ \rho$. As we shall see,
this representation is only possible when $ {\cal P} = 1$, i.e.
when there is no coarse graining at all, which means that $ l_{\rm
av}$ vanishes.
This is because eqn. (\ref{trivsol}) implies that
 the term $ \langle a| \zeta_{\hat{a}}$ in $ E_{\cal P}$
must look like
\begin{equation}
\langle a| \zeta_{\hat{a}} = \int {\rm d}1 \,\rho^{1/2} 
\epsilon'[\lambda, \varphi, \kappa, \mu, \hat{a}(x;0)] \,\rho^{1/2}\,.
\end{equation}
But $ \epsilon'$ is {\it not} independent of $ \rho$, because it
is a function of the irrelevant part of the field operator at initial
time and in general, a change of the initial state will change the values
of the relevant variables. Only in the case of $ {\cal P} = 1$, 
i.e. $ l_{\rm av} = 0$, one can overcome this dependence for an
arbitrary initial state.
In order to seek for possible other solutions of (\ref{lvar}),
we rewrite this equation in the following way
\begin{eqnarray}
\label{Ll}
\nonumber
\int {\rm d}1 \,\left(\frac{\partial \rho}{\partial l} \right)
\frac{\delta}{\delta \rho} (E_{\rm qm} + E_{\cal P}) + 
\int {\rm d}1 \,\rho \frac{\partial }{
\partial l}\,
 \frac{\delta}{\delta\rho} (E_{\rm qm} + E_{\cal P}) =
\frac{\partial}{\partial l}(E_{\rm qm} + E_{\cal P}) \,,
\end{eqnarray}
and in this form, it tells us already how $ l_{\rm av}$ has to be in order
to satisfy  equation (\ref{lvar}) resp. (\ref{Ll}). Namely,
if there is an $ l_{\rm av}$ such that 
\begin{equation}
\label{cond2}
\left[\frac{\rm \partial}{\partial l} \rho\right]_{l = l_{\rm av}} = 0 \,,
\end{equation}
then the above equation becomes valid, when at the same time
$ E_{\rm qm} + E_{\cal P}$ satisfies
\begin{equation}
\label{cond2b}
\left[\frac{\rm \partial}{\partial l} (E_{\rm qm} + E_{\cal P})
\right]_{l = l_{\rm av}} = 0
\end{equation} 
and since $ E_{\rm qm}$ does only depend on $ \rho$ it follows
\begin{equation}
\label{cond2.1b}
\left[\frac{\rm \partial}{\partial l} E_{\cal P}
\right]_{l = l_{\rm av}} = 0
\end{equation}
Furthermore, because of the consistency with the 'continuity' equation
(\ref{eqnm}), the validity of (\ref{cond2}) and (\ref{cond2b}) implies that
it must even be
\begin{equation}
\label{cond2c}
\left[\frac{\rm \partial \sigma}{\partial l}
\right]_{l = l_{\rm av}} = 0 \,,\quad \sigma = \{\varphi, \lambda,
\kappa, \mu \}\,.
\end{equation} 
Condition (\ref{cond2}) states that if we increase
our averaging length scale being in the neighborhood of $ l_{\rm av}$,
then the values of $ \lambda, \rho, \varphi, \mu$ and $
\kappa$ do not change at all, or in other words, the smallest
lenght-scale on which
the functions $ \lambda, \rho, \varphi, \mu$ and $
\kappa$ significantly vary, i.e. $ l_{\rm obs}$, is much bigger than 
the regarded scale $ l_{\rm av}$
 \footnote{The other obvious solution is when $ l_{\rm av}= \infty$.
But in this case the relevant variables become constant functions
in space and in time and hence decribes a completely homogeneized 
system with no dynamics}.
This is a remarkable result, because
it shows explicitely that it must be $l_{\rm av} \ll l_{\rm obs}$.
Since $ E_{\cal P}$ involves the coupling factor $ l_{\rm av}^4/16$,,
c.f. equation (\ref{GPapprox}) and (\ref{FPapprox}),
it follows that microscopic dissipation and fluctuations are coupled
very weakly to the macroscopic evolution.\\
However, we should stress out, that the actual existence
of a critical value $ l_{\rm av}$ has not been proven, rather we
showed that its existence would be consistent with the results
obtained from our variational approach.
\subsubsection{Nodal regions}
It is well known, that $ \delta_\rho E_{\rm qm}$
becomes singular whenever $ a(1,...,N;t)$ is on a nodal region, i.e.
a region where the wave function vanishes. In this situation,
equation (\ref{Nzwanzig}) is not consistent with equation
(\ref{euler1}) and additional 'quantization rules' are necessary to
re-establish consistency \cite{w94}. In the classical regime, of
course, we do not want to fulfill any additional constraints which
are of pure quantum nature. To do so, we want the amplitude $
\phi(1,...,N;t)$ to be strictly positive, that means, $ a(1,...,N;t)$
must not have any nodal regions. This leads us to  the following 
assumption: For a fixed $ t$ and a given $ l_{\rm av}$ 
let $ \Omega_t^{\cal P} \subset  {\sf
R}^3$ be a set defined by the relation
\begin{equation}
\label{cond1}
{\cal P}\hat{a}(x;t) = 0 \quad\mbox{iff}\quad  x \in \Omega_t^{\cal P},
\end{equation}
then there is a $ l'_{\rm av} > l_{\rm av}$ such that $ \Omega_t^{\cal
P} = \emptyset$ for all $ t$. In case of its existence, we take the
smallest value $ l'_{\rm av}$ to be the actual averaging length scale.
Obviously, this choice guaranties that the relevant part
of $ \hat{a}(x;t)$, that is all $ \hat{a}$ satisfying $ {\cal P}\hat{a} =
\hat{a}$, never vanishes. Consequently, the amplitude of the relevant $N$-particle
wavefunction, $ \phi(1,...,N;t)$, does not vanish, either.
This 'prohibition' of singularities in $ \delta_\rho E_{\rm qm}$
sets a lower bound for $ l_{\rm av}$. It can be viewed as
a condition for the system being not too far from equilibrium, thus
the locally averaged particle number density, $\rho$, is positive
everywhere.
\subsubsection{Quantum potential}
Because the quantum potential $ \delta_\rho E_{\rm qm}$ is defined
in configuration space, it reveals an inherent
non-local character in position space: variation of the density of one
particle can instantly change the motion of another particle over
an arbitrary spatial distance. This term is a product of quantum
coherence and describes a completely non-classical
phenomenon. Thus for the emergence of classicality, we want to 
give a natural bound for the quantum potential $ \delta_\rho E_{\rm qm}$ such that 
its influence remains minor at the macroscopic level of description.\\
Recall that for a given set $ {\cal O}(x,l') =
[x-l_{\rm av}/2, x + l_{\rm av}/2]^3 \subset {\sf R}^3$ and an
arbitraray $ x$ it is
\begin{equation}
\int\limits_{{\cal O}(x,l_{\rm av})} {\rm d}x' \,\rho \,\nabla \frac{\delta}{\delta \rho} E_{\rm
qm}  \rightarrow 0\,, \quad \mbox{as} \quad l_{\rm av} \rightarrow
\infty\,,
\end{equation} 
which means that by increasing $ l_{\rm av}$, the value of
$ {\cal P}\rho \nabla \delta_\rho E_{\rm qm}$ decreases, see \cite{gd82}.
Therefore, we can give a natural upper bound
for the quantum force by relating it to the disspative and stochastic
term via
\begin{equation}
\label{cond3}
\left|\left[\rho \nabla \frac{\delta}{\delta \rho} E_{\rm qm}
\right]_{l = l_{\rm av}}\right| \leq
\left|\left[ \left(\rho \nabla \frac{\kappa}{2 \rho}\frac{\delta}{\delta\kappa}
 + \nabla\rho \frac{\delta}{\delta\rho} - \sum_{\sigma \in
I \backslash \{\kappa\}} (\nabla\sigma) \frac{\delta}{\delta\sigma}
\right) E_{\cal P}\right]_{l = l_{\rm av}}\right| \,.
\end{equation}
This inequality demands for a lower bound of $ l_{\rm av}$.
Thus if $ l_{\rm av}$ turns out to be  consistent with that bound, then relation
(\ref{cond3}) tells us that
'quantum forces' never exceed those force terms coming from
microspcopic dissipation and fluctuation. The latter in turn are only
weakly coupled to the macroscopic system , because $ E_{\cal P}$ involves 
the factor $ l_{\rm av}^4/16$.
In particular, the above condition for a proper choice of the averaging length scale
reveals a rather delicate interaction between the
suppression of quantum coherence and classical
determinism of the equation of motion. This is because to accomplish
the former one has to increase $ l_{\rm av}$ in order to deminish the influence
of the quantum potential. But on the other hand, the same mechanism
increases the strength of the fluctuation term, which possibly leads
to a highly stochastic, i.e. non-deterministic, dynamical evolution.\\
Hence, for the equations of motion (\ref{eqnm}) - (\ref{eqnme}), the classical realm is
entered if $ l_{\rm av}$ is large enough such that the local density 
is positive and if its value  additionally implies that 
equation (\ref{cond2b}) is consistent with (\ref{cond3}). But at the same time it must also 
comply with $ l_{\rm av} \ll l_{\rm obs}$, which is in agreement with 
equation (\ref{Ll}) and (\ref{cond2c}). Then -- on macroscopical scales -- these equations turn out to be of
the same form as the hydrodynamical equations. They are conservation
laws representing mass continuity, vortex dynamics and momentum flow. 

\subsection{Temperature}

Apart from mass and momentum density, classical  macroscopic systems
at statistical equilibrium are additionally characterized by their
temperature. 
Thus any model that asserts to describe classical systems adequately
must also give a description of this particluar physical term. 
In the previous section we showed that averaging length could 
be vanishing or infinite. However, if a $ l_{\rm av}$
exists inbetween those bounds such that it fulfills the constraints
(\ref{cond2}) -  (\ref{cond2c}) then the emergent system is deterministic and local in
time. In this situation the system is in a local equilibrium, because
the average over a volume of size $ \sim l_{\rm obs}$ of an aribtrary
macroscopic function is the same when averaged over a smaller part
of this volume. This holds obviuosly because any macrsoscopic function
becomes constant in space on spatial scales smaller than $ l_{\rm obs}$.
In this case it should be possible to assign a certain temperature
to the system. This assignment is quite clear for the extreme cases, 
when $ l_{\rm av}$ is either infinite or zero: For the latter case there is 
no coarse grained system that gives rise to an extremum of the 
action $ S = \int L \,{\rm d}t$. In this situation the system is described
by the original theory and from the classical point of view the
corresponding temperature is infinite. On the other hand, when
$ l_{\rm av} = \infty$ is consistent with the equations of motion and
the constraints coming from the Hamilitonian principle then the system
is described completely by constant functions in scpace and
time. This case corresponds to a zero temperature scenario where there
are no fluctuations and inhomgeneities. Of course, these two cases 
represent the {\it trivial critical points} of infinite and zero temperature. 
In any other situation, this is where $ l_{\rm av}$ turns out to be positive and finite,
the system's temperature $ T$ should be a function of $ l_{\rm av}$ and
possibly of all other relevant parameters (such as the particle mass $
m$) of the original theory.
If these 'other' parameters remain fixed then $ l_{\rm av}$ can
actually be identified with the temperature $ T$. For non-interacting 
and non-relativistic fields this relation should read as
\begin{equation}
\frac{\hbar^2}{2 \,m\, l^2_{\rm av}} = k_{\rm B} T \,,
\end{equation}
where $ k_{\rm B}$ is Boltzmann's constant.\\
Thus temperature can only be assigned to the macroscopic system,
when the above mentioned conditions hold and from which it follows that
the system is in a local equilibrium. In general however, the averaging
length $ l_{\rm av}$ will not have anything to do with a physical
temperature. This situation is very similar to the role of temperature in non-equilibrium statistical mechanics, where the parameter
$ T$ appearing in the distribution function turns out to be the temperature
only if the system described by this distribtution function
is in (local) equilibrium.

\section{Conclusions and remarks}

In this work, we show that it is possible to study 
the emergence of classical dynamics of systems with many degrees of
freedom directly by applying the coarse graining principle to
the quantum equations of motion. This lead us to a evolution 
equation for the relevant part of the $ N$-particle wave function.
Out of this we are able to construct a Lagrangian functional $ L$.
This functional is  expressed by means of local densities that are defined
through the kinetic part of the Lagrangian.
Variation of $ L$ then leads us to a sample of dynamic equations which
explicitely contain terms arising from dissipation, fluctuation and quantum
coherence.  This approach
is somewhat similar to the deduction of the Madelung equations (or
{\it quantum fluid dynamic} equations}, henceforth QFD equations) from
 the  Schr\"odinger
equation. However, there are formal and conceptual distinctions
between the latter and our approach. Equation (\ref{Nzwanzig}) is 
equal to the $ N$-particle Schr\"odinger equation only in case of
$ l_{\rm av} = 0$. For $ l_{\rm av} > 0$ it is {\it
non-linear} in the relevant part of the wavefunction, because the noise term
$ \zeta_{\hat{a}}$ breaks the original linearity. Moreover, we obtain
non-local bahavior in time due to the dissipative term $ {\cal
G^P} a$. All these features are not present in ordinary
derivations of the QFD equations.
The most prominent difference is perhaps that QFD equations do
not indicate how a classical limit could ever be achieved.
Only by carefully applying the coarse graining principle we 
were able to describe this limit and the transition towards it.\\
The classical limit itself is -- as we believe -- characterized by the
following 
properties: locality in space, locality in time and determinism.
To accomplish the demand for determinsm and locality in time, we
vary the Lagrangian $ L$ also with respect to $ l_{\rm av}$, which 
is a generalization of the variational approach presented in
section III. A.
Then we show that the resulting constraints becomes valid when a
critical value of $ l_{\rm av}$ exists being consistent with the
conditions for classicality stated above.  

To avoid non-local quantum behavior in space we additionally need 
to show that the influece of the quantum potential 
$ \delta_\rho E_{\rm av}$ becomes neglegible. This happens
when the quantum potential does not exceed a bound given
by the disspative and fluctuation force terms,
c.f. condition (\ref{cond3}). Additionally, we have
to demand that there are no nodal regions in the coarse
grained wavefunctions. Then, provided such a  
value of $ l_{\rm av}$ exists, the equations of motion become
classical. In our case, we recover the Euler equation
together with the equations for continuity and for the motion
of vortex lines.

We think that our approach to the problem of emergent classicality
is quite general, because it requires only basic principles.
Starting from a quantum field theory one additionally introduces
a mininum level of spatial resolution. Then the immediate question
arises how the original theory looks like when is represented
on an arbitrary level of coarse graining. Finally, 
locality in space/time and determinism -- as principles of the
classical realm -- are questioned and eventually the 
coarse grained equations of motion  tell us whether they
are consistent with these principles or not. 

This approach should also work for interacting fields
as long as not too high energy densities are considered.
In general, interacting fields do not posses a representation
in a fixed $ N$-particle Hilbertspace anymore. This is because of
the creation and annihilation of particles. Thus if $ N$ varies
strongly in time the Lagrangian functional approach presented here 
breaks down completely
and our description of emergent classicality becomes inadmissable.
In this situation we expect the original (quantum field) theory 
to be an appropriate model of physical description.

As a final remark we want to point out that there is another
parameter in the Lagrangian functional $ L$ which has not been
considered explicitely, yet. This is the number of spatial dimensions
$ d$. When analytically continued along the real line, it could
be used as an extra variable in the variational scheme presented in 
this work. The outcome of a corresponding constraint could then
give some further insight to questions like: Does the existence of a classical
domain require a certain number of spatial dimensions?
The hope is that this number is '3' in the low-energy limit of
the known interactions.


%

%

%
%



\begin{references}
\bibitem{gd82} S.K. Ghosh and B.M. Deb, Phys. Rep. {\bf 92}, 1
(1982)
\bibitem{w94} T.C. Wallstrom, Phys. Rev. A  {\bf 49}, 1613
(1994)
\bibitem{gmh93} M. Gell-Mann and J.B Hartle, Phys. Rev. D  {\bf 47}, 3345
(1993)
\bibitem{hs78} G. Holzwarth and D. Sch\"utte, Phys. Lett. B  {\bf 73}, 255
(1978)
\bibitem{a98} C. Anastopoulos, preprint gr-qc/9805074 (1998)
\bibitem{rm96} J. Rau and B. M\"uller, Phys. Rep. {\bf 272}, 1
(1996)
\bibitem{t53} A. Thellung, Physica {\bf 19}, 217
(1953)
\end{references}
\end{document}